\documentclass[useAMS,usenatbib]{rmaa1}

\begin{document}

\title{An approach to effective temperature and surface gravity in post-AGB and RV Tauri stars 
at the near-IR region}

\author{R. E. Molina,\altaffilmark{1}
\medskip }

\altaffiltext{1}{Laboratorio de Investigaci\'on en F\1sica Aplicada y Computacional,
Universidad Nacional Experimental del T\'achira, Venezuela.}

\fulladdresses{
\item R. E. Molina: Laboratorio de Investigaci\'on en F\1sica Aplicada y Computacional,
Universidad Nacional Experimental del T\'achira,  Venezuela,
(rmolina@unet.edu.ve).}

\shortauthor{Molina}
\shorttitle{T$_{\rm eff}$ and log~$g$ at the near-IR region}

\resumen{Se determina la temperatura efectiva y la gravedad superficial de 
un conjunto de estrellas
evolucionadas post-AGB y RV Tau a partir de ecuaciones emp\1ricas usando como
calibradores los colores intr\1nsecos de la fotometr\1a 2MASS (Two Micron All Sky Survey).
Analizamos un total de 36 estrellas donde 25 de ellas son post-AGB y las restantes 11 son RV Tau 
respectivamente. Utilizamos un grupo de 11 estrellas con medidas de paralaje como
calibradores en la determinaci\'on de la magnitud absoluta en la banda J.
Los resultados para la T$_{\rm eff}$ y la log~$g$ derivadas de los colores intr\1nsecos
(J--H)$_{0}$ y (H--K$_{s}$)$_{0}$ alcanzan una dispersi\'on de 220 K y 0.27, respectivamente. 
Estimamos la magnitud absoluta en la banda J con una
precisi\'on de 0.28 mag, esto nos indica que los colores (J--H)$_{0}$ y (H--K$_{s}$)$_{0}$
presentan sensibilidad a dicho par\'ametro.}

\abstract{A number of empirical correlations that allows us to calculate the effective
temperature and surface gravity for a set evolved post-AGB and RV Tauri stars are
determined using as calibrators the intrinsic colours of 2MASS (Two Micron All Sky Survey)
photometry. We have analyzed a total sample of 36 stars where 25 are post-AGB stars and
11 are RV Tauri stars, respectively. A group of 11 stars with parallaxes measures
were used as calibrators of the absolute magnitude.
The result for T$_{\rm eff}$ and log~g from intrinsic colours (J--H)$_{0}$ and
(H--K$_{s}$)$_{0}$ at the near-infrared pass bands reach a dispersion of 220 K and 0.27, 
respectively. We can estimate the absolute magnitude using the
intrinsic colour in the near-infrared band with an uncertainty of 0.28 mag.
This indicates that (J--H)$_{0}$ and (H--K$_{s}$)$_{0}$ show sensitivity
to the absolute magnitude.}

\keywords{Fundamental parameters: T$_{\rm eff}$, log~$g$ and $M_{V}$;
stars: post-AGB--RV Tauri stars.}

\maketitle

\section{INTRODUCTION}
\label{sec:introd}

Post-AGB stars are low and intermediate mass stars (0.8 -- 8 M$_{\odot}$)
that have passed the asymtotic giant branch (AGB) and are on their way toward
the planetary nebula (PN) phase. On the other hand, RV Tauri stars are
very luminous and variable stars ($M_{V}$ $\approx$ $-$3 mag)
with spectral types F, G and K which have an infrared excess attributed to
the mass lost in the AGB phase (Preston 1963; Jura 1986; Wahlgren 1993).
According to RV Tauri stars' infrared and statistical properties,
Jura (1986) suggests that the RV Tauri stars are also post-AGB stars.
The determination of fundamental physical parameters for these types of objects
(post-AGB and RV Tauri)
such as effective temperature and surface gravity are obtained from
the ionization balance for elements with two states of
ionization (e.g. the ratio of iron Fe I/Fe II) using atmospheric models
for this purpose (see Giridhar et al. 2010 paper I).

Alternatively, attempts have been made to derive a set of functional
relationships that allow to estimate these parameters from correlations with
photometric data for these objects. The proposed first attempts were
carried out by Arellano Ferro et al. (1990), Arellano Ferro y Mantegazza (1996)
using the reddening-free indexes [$c_1$], [$m_1$] and [$u-v$] from Str\"omgren photometry.
Later, Arellano Ferro (personal communication) found functional relationships
involving the effective temperature and photometric indexes [$c_1$], [$m_1$] and
[u-v] using 41 bright supergiant stars with classes I and II with spectral types
within the range of A0--K0 from Bravo-Alfaro et al. (1997). However, these
relations were never published since their intrinsic dispersion was large,
probably due to the limited quality of used temperature data.

Recently, Arellano Ferro (2010) has estimated functional relationships
from a set of 50
supergiant stars with spectral types F--G including some objects like stars
evolved post-AGB and RV Tauri stars (see Stasi\'nska et al. 2006). For calibration,
A. Ferro uses data from a homogeneous sample of temperature and gravity
taken from Lyubimkov et al. (2010). This author concludes that the temperature
can be obtained with good accuracy from the reddening-free indexes [$c_1$]
and  [$m_1$]  (e.g.$\pm$ 152 K for both indexes) while gravity can be calculated
from index $\Delta$[$c_1$] with an uncertainty of 0.26 dex. However,
although there is a correlation between the absolute magnitude $M_{V}$ and the
index $\Delta$[$c_1$], this correlation is less satisfactory.

In the near-infrared region, 2MASS catalogue (Cutri et al. 2003) provides the
most complete database of near infrared Galactic point sources available to date.
Various jobs in this region have shown the dependence of atmospheric parameters
with the de-reddened colours (Bessell \& Brett 1988; Covey et al. 2007; Bilir et al. 2008a;
2008b; 2009; Straizys \& Lazauskaite 2009; Yaz et al.2010). Bilir et al. (2008a)
obtained the transformation formulae between JHK$_{s}$ photometry system to BVRI and SDSS gri
photometric systems for the main sequence stars. 
These transformations provide absolute magnitude and distance determinations which can
be used in space density evaluations at short distances were some or all of the gri
magnitudes are saturated. These authors also showed that those
formulae were sensitive to metallicity. In other similar studies (Bilir et al.
2008b; Bilir et al. 2009) the calculated absolute magnitudes were compared with
Pickles (1998)' synthetic data and the results were in good agreement with
synthetic library results. In recent study (Yaz et al. 2010) it was obtained the
transformation formulae for the conversion between JHK$_{s}$ photometry system to BVRI, 
and SDSS gri photometric systems for set of giant stars. 
These formulae are directly dependent on metallicity as well and also provide absolute 
magnitude and distance determinations for this type of population.

In this paper we introduce for the first time two functional relationships that
allow us to estimate the effective temperature and gravity in post-AGB and RV
Tauri stars from their infrared colours. These equations are valid for evolved
stars of low and intermediate masses (1 $<$ M$_{\odot}$ $\leq$ 4) in which the central 
star is visible due
to the low opacity of the dust that surround them. However, most of these
objects radiate in the infrared range and are, therefore, detectables by
2MASS source. Furthermore, these equations can not be applied to massive
post-AGB objects, i.e., 4 $<$ M$_{\odot}$ $\leq$ 8, without optical counterpart owing to 
a dense circumstellar
material that obscures the central star. More massive objects will have more material
in their circumstellar shells and the star will be quite obscured, while it will not
be in the case of objects will less material in the envelope. Sources with dusty shells
will therefore be fainter at optical wavelenghts but brighter in the infrared. 
Inhomogeneous sample from several authors and the compilation of Stasi\'nska et al.
(2006) have been used as calibrators. We attempt to recover the
absolute magnitude from the intrinsic colour
despite the lack of homogeneity and the limitations on the parallax of
the sample.

The structure of the paper is as follows: Sect. 2 introduces the
data used and the criteria applied to estimate the colour excess.
Sect. 3 shows the funcional relationships for atmospheric parameters
using intrinsic colours. Sect. 4 deals with a relationship to
estimate the absolute magnitude in the J passband using as calibrators
the available Hipparcos trigonometric parallax measurements.
Finally, in Sect. 5 we discuss our results.

\begin{figure}
\begin{center}
\includegraphics[width=7.5cm,height=7.5cm]{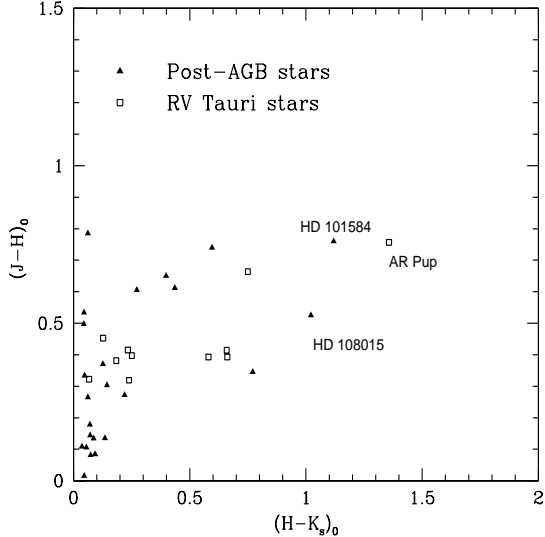}
\caption{The J-H vs. H-K$_{s}$ diagram for unreddened all stars of the sample.
The filled triangles represent the post-AGB stars and open squares to
the RV Tau stars.}
    \label{fig1}
\end{center}
\end{figure}

\section{Data}
\label{sec:data}

For post-AGB stars the temperature and gravity used in our calibration come mainly from
the compilation made by Stasi\'nska et al. (2006) while for RV Tauri stars
are obtained from Giridhar et al. (2005). We select the atmospheric parameters for a set of
24 post-AGB stars and 11 RV Tauri stars, respectively (see column 8 of table~\ref{table1}).
A group of 11 stars with Hipparcos parallax measurements were used as calibrators of the absolute
magnitude (see Table~\ref{table3}). 
Most stars were selected with intermediate and high galactic latitude to reduce the 
contribution of the observed extinction. Stars identified on the galactic plane were 
chosen to have parallax measurements.
Figure~\ref{fig1} shows the unreddened colour-colour diagram of
the post-AGB (filled triangles) and RV Tauri (open squares) stars. We notice that
there is a good correlation between the sample with a large dispersion probably due
to the dependence on metallicity. Some objects in the sample show a greatest colour
(H--K$_{s}$) $>$ 1.0, e.g. AR Pup, HD 101584 and HD 108015. 
This excess seen in the (H--K$_{s}$) colour may be
due to the contribution of the circumstellar component since the contribution of interstellar
dust has been removed (for details see Section~\ref{sec:redden}).

\begin{figure}
\begin{center}
\includegraphics[width=7.5cm,height=7.5cm]{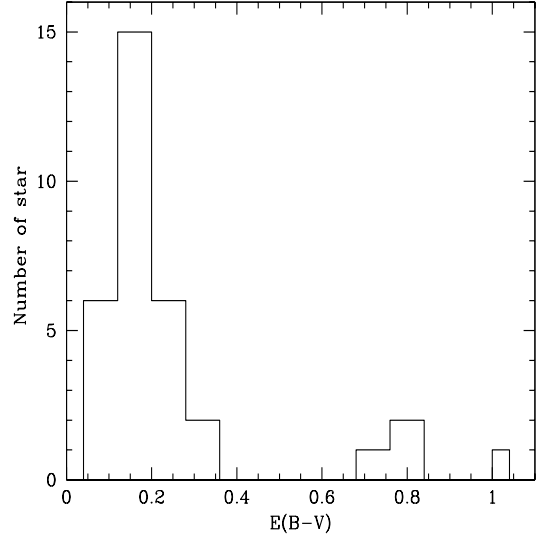}
\caption{The distribution of the total E(B--V) found for our sample.}
    \label{fig2}
\end{center}
\end{figure}

We studied the post-AGB and RV Tauri stars with A--G spectral type within the range in effective
temperature and gravity between 5000--8500 K and 0.0--2.0,
respectively. According to SIMBAD\footnote{http://simbad.u-strasbg.fr/simbad/} source a few
post-AGB stars in the sample are considered semi-regular pulsanting stars, while the RV Tauri
are mostly variable stars with exception of HP Lyr (eclipsing binary star) and UY CMa
(semi-regular pulsanting star).
These objects are indicated by the upper subscripts ''a", ''b" and ''c".
Table~\ref{table1} contains the fundamental atmospheric
parameters for the total number of stars studied. Columns are organized according to the
type of star and the HD number as follows:
name of stars, spectral type, galactic coordinates, effective temperature, surface gravity,
comment and references.

The trigonometric parallaxes used in our 11 stars were taken from the current reduced Hipparcos
catalogue (van Leeuwen 2007). The JHK$_{s}$ photometry was taken from 2MASS
(Cutri et al. 2003) source.

\begin{table*}
 \centering
 \begin{minipage}{140mm}
  \caption{The fundamental atmospheric parameters for the sample post-AGB and
RV Tauri used as calibrators. The sources of temperature and gravity are
respectively by the numbers in column 8.}
 \label{table1}
\begin{tabular}{lllrrllc}
  \hline
\multicolumn{1}{l}{Star}&
\multicolumn{1}{l}{SpT.}&
\multicolumn{1}{c}{{\it l}}&
\multicolumn{1}{c}{{\it b}}&
\multicolumn{1}{c}{T$_{\rm eff}$}&
\multicolumn{1}{l}{log~g}&
\multicolumn{1}{l}{comment}&
\multicolumn{1}{l}{Refs.}\\
\multicolumn{1}{l}{}&
\multicolumn{1}{l}{}&
\multicolumn{1}{c}{($^{o}$)}&
\multicolumn{1}{c}{($^{o}$)}&
\multicolumn{1}{c}{(K)}&
\multicolumn{1}{l}{}&
\multicolumn{1}{l}{}&
\multicolumn{1}{l}{}\\

 \hline
HD 27381      & F2          & 161.63  & $-08.07$ & 7500$\pm$250 & 1.0$\pm$0.5   & post-AGB & (01)\\
HD 46703      & F7IVw       & 161.98  & $+19.59$ & 6000$\pm$150 & 0.4$\pm$0.3   & post-AGB$^{a}$ & (02)\\
HD 56126      & F5Iab       & 206.74  & $+09.99$ & 7250$\pm$200 & 0.5$\pm$0.5   & post-AGB$^{a}$ & (16)\\
HD 95767      & F3II        & 290.53  & $-01.95$ & 7300$\pm$300 & 1.5$\pm$0.25  & post-AGB & (11)\\
HD 101584     & F0Iap       & 293.02  & $+05.93$ & 8500$\pm$500 & 1.5$\pm$0.5   & post-AGB & (03)\\
HD 107369     & A2II/III    & 295.47  & $+29.86$ & 7600$\pm$200 & 1.5$\pm$0.25  & post-AGB & (11)\\
HD 108015     & F4Ib/II     & 298.25  & $+15.47$ & 6800$\pm$200 & 1.25$\pm$0.25 & post-AGB$^{a}$ & (11)\\
HD 112374     & F3Ia        & 304.34  & $+36.39$ & 6000$\pm$275 & 0.6$\pm$0.3   & post-AGB$^{a}$ & (12)\\
HD 114855     & F5Ia/Iab    & 306.24  & $+08.03$ & 6000$\pm$200 & 0.5$\pm$0.25  & post-AGB & (10)\\
HD 116745     & F0Ibp       & 309.07  & $+15.17$ & 6950$\pm$75  & 1.15$\pm$0.1  & post-AGB & (13)\\
HD 133656     & A1/A2Ib/II  & 325.03  & $+08.64$ & 8000$\pm$200 & 1.25$\pm$0.25 & post-AGB & (14)\\
HD 148743     & A7Ib        & 007.95  & $+26.70$ & 7200$\pm$500 & 0.5$\pm$0.3   & post-AGB & (04)\\
HD 161796     & F3Ib        & 077.13  & $+30.86$ & 6666$\pm$500 & 0.7$\pm$0.3   & post-AGB$^{a}$ & (08)\\
HD 163506     & F2Ib        & 051.53  & $+23.28$ & 6550$\pm$500 & 0.6$\pm$0.3   & post-AGB$^{a}$ & (02)\\
HD 172481     & F2Ia        & 006.72  & $-10.73$ & 7250$\pm$200 & 1.5$\pm$0.25  & post-AGB$^{a}$ & (15)\\
HD 179821     & G5Ia        & 035.61  & $-04.95$ & 6800$\pm$150 & 1.3$\pm$0.5   & post-AGB$^{a}$ & (05)\\
HD 190390     & F1III       & 030.59  & $-21.53$ & 6600$\pm$500 & 1.6$\pm$0.64  & post-AGB$^{a}$ & (04)\\
CpD-625428    & A7Iab       & 326.20  & $-11.11$ & 7250$\pm$200 & 0.5$\pm$0.25  & post-AGB & (10)\\
IRAS05113+1347& G8Ia        & 188.85  & $-14.29$ & 5250$\pm$150 & 0.25$\pm$0.5  & post-AGB & (17)\\
IRAS05341+0852& F4Iab:      & 196.18  & $-12.14$ & 6500$\pm$250 & 1.0$\pm$0.5   & post-AGB & (18)\\
IRAS07430+1115& G5Ia        & 208.93  & $+17.06$ & 6000$\pm$250 & 1.0$\pm$0.25  & post-AGB & (21)\\
IRAS18095+2704& F3Ib        & 053.83  & $+20.18$ & 6500$\pm$150 & 0.5$\pm$0.5   & post-AGB & (20)\\
IRAS19386+0155& F           & 040.50  & $-10.08$ & 6800$\pm$100 & 1.4$\pm$0.2   & post-AGB & (19)\\
IRAS22223+4327& F9Ia        & 096.75  & $-11.55$ & 6500$\pm$350 & 1.0$\pm$0.3   & post-AGB$^{a}$ & (22)\\
\hline
HD 82084      & A4Ib/II     & 282.42  & $-09.23$ & 6700$\pm$200 & 2.0$\pm$0.2   & RV Tauri$^{c}$ & (23)\\
HD 105578     & F7/F8Ib     & 295.24  & $+16.81$ & 6000$\pm$250 & 1.0$\pm$0.5   & RV Tauri$^{c}$ & (06)\\
HD 107439     & G4Vp        & 297.87  & $+13.36$ & 6250$\pm$250 & 1.25$\pm$0.5  & RV Tauri$^{c}$ & (06)\\
HD 170756     & K0III       & 028.49  & $-03.77$ & 5900$\pm$150 & 1.13$\pm$0.15 & RV Tauri$^{c}$ & (07)\\
AR Pup        & F0Iab       & 253.02  & $-02.99$ & 6300$\pm$200 & 1.5$\pm$0.2   & RV Tauri$^{c}$ & (09)\\
AR Sgr        & G4          & 012.39  & $-12.28$ & 5300$\pm$200 & 0.5$\pm$0.2   & RV Tauri$^{c}$ & (24)\\
HP Lyr        & A6          & 072.01  & $+11.71$ & 6300$\pm$200 & 1.0$\pm$0.2   & RV Tauri$^{b}$ & (24)\\
RX Cap        & G2          & 030.35  & $-24.27$ & 5800$\pm$200 & 1.0$\pm$0.2   & RV Tauri$^{c}$ & (24)\\
TX Oph        & G0          & 024.74  & $+26.11$ & 5000$\pm$200 & 0.5$\pm$0.2   & RV Tauri$^{c}$ & (24)\\
UY CMa        & G0V         & 224.75  & $-14.85$ & 5500$\pm$200 & 0.0$\pm$0.2   & RV Tauri$^{a}$ & (24)\\
UZ Oph        & G2          & 028.87  & $+23.01$ & 5000$\pm$200 & 0.5$\pm$0.2   & RV Tauri$^{c}$ & (24)\\

\hline
\end{tabular}
\end{minipage}

\scriptsize{\begin{flushleft}
$^{a}$ Semi-regular pulsanting stars.\\
$^{b}$ Eclipsing binary star.\\
$^{c}$ Variable stars of RV Tau type.\\
\end{flushleft}}

\scriptsize{\begin{flushleft}
(01) Giridhar \& Arellano Ferro  (2005), (02) Luck \& Bond (1984),
 (03) Sivarani et al. (1999), (04) Luck et al. (1990), (05) Za$\breve{c}$s et al. (1996), (06) Mass
 et al. (2002), (07) Giridhar et al. (1998), (08) Cenarro et al. (2007),
 (09) Gonz\'alez et al. (1997a),
 (10) Giridhar et al. (2010), (11) van Winckel et al. (1997), (12) Luck et al. (1983),
(13) Gonz\'alez \& Wallerstein (1992), (14) van Winckel et al. (1996), (15) Arellano Ferro et al. (2001),
(16) Hrivnak \& Reddy (2003), (17) Reddy et al. (2002), (18) van Winckel \& Reyniers (2000), (19) Pereira et al. (2004),
(20) Sahin et al. (2010), (21) Reddy et al. (1999), (22) Decin et al. (1998), (23) Giridhar et al. (1994), (24)
Giridhar et al. (2005)
\end{flushleft}} 

\end{table*}

\begin{table*}
 \centering
 \begin{minipage}{140mm}
  \caption{The total colour excess and the colour excess of circumstellar and interstellar
components for common objects in both samples.}
 \label{table2}
\begin{tabular}{rrccc}
  \hline
\multicolumn{1}{c}{IRAS/HD}&
\multicolumn{1}{c}{E(B--V)}&
\multicolumn{1}{c}{E(B--V)$_{IS}$}&
\multicolumn{1}{c}{E(B--V)$_{IS}$}&
\multicolumn{1}{c}{E(B--V)$_{CS}$}\\
\multicolumn{1}{c}{number}&
\multicolumn{1}{c}{(observed)}&
\multicolumn{1}{c}{(Luna et al.)}&
\multicolumn{1}{c}{(this work)}&
\multicolumn{1}{c}{(Luna et al.)}\\

\hline
05113+1347 & 1.1$\pm$0.2   & 0.14  & 0.310 & 0.96 \\
05341+0852 & 1.65$\pm$0.09 & 0.16  & 0.215 & 1.49 \\
07134+1005/ & 0.4$\pm$0.1   & 0.024 & 0.080 & 0.38 \\
56126      &               &       &       &      \\
17436+5003/ & 0.24$\pm$0.09 & 0.026 & 0.030 & 0.18 \\
161796     &               &       &       &      \\
19114+0002/ & 0.60$\pm$0.05 & 0.54  & 0.739 & \ldots \\
179821     &               &       &       &      \\
19386+0155 & 1.05$\pm$0.09 & 0.35  & 0.299 & 0.80 \\
22223+4327 & 0.2$\pm$0.1   & 0.21  & 0.147 & \ldots \\
\hline
\end{tabular}
\end{minipage}
\end{table*}

\begin{table*}
 \centering
 \begin{minipage}{140mm}
  \caption{JHK$_{s}$ photometry, the colour excess and intrinsic colours of the total study sample.}
\label {table3}
\begin{tabular}{lccccrc}
  \hline
\multicolumn{1}{l}{Star}&
\multicolumn{1}{c}{J}&
\multicolumn{1}{c}{H}&
\multicolumn{1}{c}{K$_{s}$}&
\multicolumn{1}{c}{E(B--V)$_{IS}$}&
\multicolumn{1}{c}{(J--H)$_{0}$}&
\multicolumn{1}{c}{(H--K$_{s}$)$_{0}$}\\
\multicolumn{1}{c}{}&
\multicolumn{1}{c}{(mag)}&
\multicolumn{1}{c}{(mag)}&
\multicolumn{1}{c}{(mag)}&
\multicolumn{1}{c}{}&
\multicolumn{1}{c}{(mag)}&
\multicolumn{1}{c}{(mag)}\\

 \hline
HD 27381       & 5.019$\pm$0.037 & 4.747$\pm$0.033 & 4.555$\pm$0.016 & 0.797 & 0.015 & 0.046\\
HD 46703       & 7.805$\pm$0.020 & 7.513$\pm$0.017 & 7.435$\pm$0.018 & 0.085 & 0.265 & 0.062 \\
HD 56126       & 6.868$\pm$0.021 & 6.708$\pm$0.036 & 6.606$\pm$0.017 & 0.080 & 0.134 & 0.087\\
HD 95767       & 7.098$\pm$0.023 & 6.532$\pm$0.040 & 5.636$\pm$0.023 & 0.685 & 0.345 & 0.771\\
HD 101584      & 5.947$\pm$0.020 & 5.138$\pm$0.026 & 3.991$\pm$0.260 & 0.150 & 0.760 & 1.119 \\
HD 107369      & 8.977$\pm$0.023 & 8.844$\pm$0.021 & 8.794$\pm$0.021 & 0.079 & 0.108 & 0.036\\
HD 108015      & 6.94$\pm$0.03   & 6.38$\pm$0.04   & 5.338$\pm$0.027 & 0.109 & 0.525 & 1.022\\
HD 112374      & 5.251$\pm$0.019 & 4.856$\pm$0.076 & 4.715$\pm$0.017 & 0.079 & 0.370 & 0.127 \\
HD 114855      & 6.514$\pm$0.026 & 6.173$\pm$0.044 & 6.007$\pm$0.021 & 0.116 & 0.303 & 0.144\\
HD 116745      & 9.698$\pm$0.022 & 9.481$\pm$0.024 & 9.388$\pm$0.020 & 0.120 & 0.178 & 0.071\\
HD 133656      & 6.680$\pm$0.018 & 6.524$\pm$0.027 & 6.408$\pm$0.020 & 0.230 & 0.082 & 0.074\\
HD 148743      & 5.511$\pm$0.019 & 5.358$\pm$0.024 & 5.226$\pm$0.023 & 0.213 & 0.084 & 0.093 \\
HD 161796      & 6.095$\pm$0.019 & 5.979$\pm$0.033 & 5.919$\pm$0.017 & 0.030 & 0.106 & 0.055\\
HD 163506      & 4.998$\pm$0.242 & 4.239$\pm$0.036 & 3.632$\pm$0.280 & 0.058 & 0.740 & 0.596 \\
HD 172481      & 6.580$\pm$0.023 & 5.877$\pm$0.034 & 5.449$\pm$0.017 & 0.166 & 0.650 & 0.398\\
HD 179821      & 5.371$\pm$0.023 & 4.998$\pm$0.023 & 4.728$\pm$0.020 & 0.739 & 0.135 & 0.135 \\
HD 190390      & 5.011$\pm$0.037 & 4.639$\pm$0.224 & 4.563$\pm$0.017 & 0.113 & 0.334 & 0.048 \\
CpD-625428     & 9.058$\pm$0.020 & 8.871$\pm$0.025 & 8.774$\pm$0.021 & 0.134 & 0.144 & 0.072\\
IRAS05113+1347 & 9.020$\pm$0.026 & 8.423$\pm$0.018 & 8.171$\pm$0.023 & 0.310 & 0.497 & 0.044\\
IRAS05341+0852 &10.009$\pm$0.023 & 9.405$\pm$0.022 & 9.108$\pm$0.021 & 0.215 & 0.534 & 0.045\\
IRAS07430+1115 & 8.836$\pm$0.021 & 8.21$\pm$0.03   & 7.766$\pm$0.024 & 0.043 & 0.612 & 0.436\\
IRAS18095+2704 & 7.366$\pm$0.026 & 6.728$\pm$0.017 & 6.438$\pm$0.021 & 0.099 & 0.606 & 0.272\\
IRAS19386+0155 & 7.951$\pm$0.029 & 7.069$\pm$0.033 & 6.011$\pm$0.023 & 0.299 & 0.785 & 0.062\\
IRAS22223+4327 & 7.812$\pm$0.018 & 7.493$\pm$0.024 & 7.246$\pm$0.016 & 0.147 & 0.272 & 0.220\\
\hline
HD 82084       & 5.875$\pm$0.019 & 5.151$\pm$0.015 & 4.367$\pm$0.256 & 0.185 & 0.664 & 0.750\\
HD 105578      & 7.616$\pm$0.026 & 7.186$\pm$0.042 & 6.917$\pm$0.020 & 0.098 & 0.397 & 0.251 \\
HD 107439      & 7.875$\pm$0.019 & 7.443$\pm$0.038 & 6.759$\pm$0.016 & 0.122 & 0.393 & 0.662\\
HD 170756      & 5.700$\pm$0.018 & 5.338$\pm$0.017 & 5.075$\pm$0.016 & 0.128 & 0.319 & 0.239 \\
AR Pup         & 7.891$\pm$0.029 & 6.82$\pm$0.04   & 5.285$\pm$0.020 & 0.974 & 0.757 & 1.357 \\
AR Sgr         & 8.412$\pm$0.029 & 7.948$\pm$0.044 & 7.677$\pm$0.029 & 0.153 & 0.415 & 0.234\\
HP Lyr         & 8.877$\pm$0.020 & 8.429$\pm$0.020 & 7.750$\pm$0.024 & 0.107 & 0.414 & 0.659\\
RX Cap         & 9.818$\pm$0.021 & 9.463$\pm$0.025 & 9.377$\pm$0.021 & 0.103 & 0.322 & 0.067\\
TX Oph         & 8.266$\pm$0.029 & 7.846$\pm$0.059 & 7.640$\pm$0.018 & 0.121 & 0.381 & 0.184\\
UY CMa         & 9.042$\pm$0.027 & 8.597$\pm$0.057 & 7.986$\pm$0.026 & 0.163 & 0.393 & 0.581\\
UZ Oph         & 8.43$\pm$0.05   & 7.94$\pm$0.05   & 7.791$\pm$0.023 & 0.114 & 0.453 & 0.128\\

\hline
\end{tabular}
\end{minipage}
\end{table*}

\section{Reddening estimation}
\label{sec:redden}

Light coming from a star is attenuated and reddened by material in the line of
sight. Note that the total line of sight extinction is likely to contain both a
circumstellar and an interstellar component.
In order to perform our calibrations we need to de-redden the colour indexes in
the near-infrared passband. Unfortunately, there is no precise measurements in the
literature about the colour excess for each object. To estimate the colour excess
E(B--V) we use the NASA Extragalactic Database (NED, based on the work of Schlegel
et al. 1998).\footnote{http:
//nedwww.ipac.caltech.edu/forms/calculator.htlm}

\begin{figure}
\begin{center}
\includegraphics[width=7.5cm,height=7.5cm]{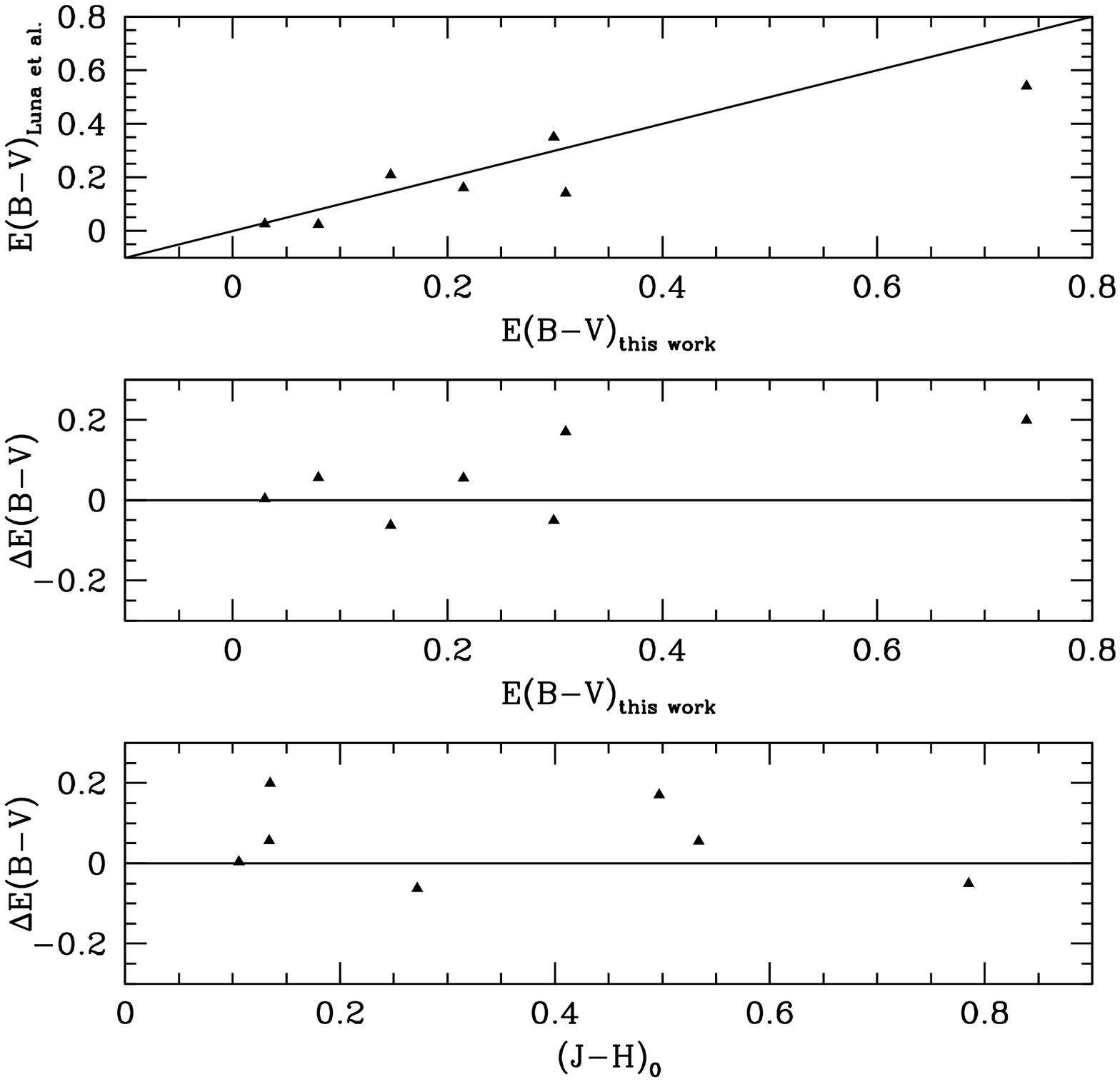}
\caption{Representation of the correlation between the colour excess estimated from the
maps by Drimmel et al. (2003) and Schlegel et al. (1998) for 7 common objects in 
both samples (upper panel). The straight line represent a slope to 45 degree. The
middle and bottom panel show of E(B--V) residuals as a function of E(B--V) derived
by Schlegel et al.'s map and the intrinsic colour (J--H)$_{0}$.}
    \label{fig3}
\end{center}
\end{figure}

This model provides E(B--V) from the galactic coordinates with an accuracy of 10\%
for stars with galactic latitudes {\it b} $>$ 5$^{0}$. For lower latitudes their reddenings are
more uncertain. The reddening estimated from NED 
refers to the full line of sight and should be applied as it stands only to extragalactic
objects or objects well about the dust layer.
Several studies have noticed that the dust model of Schlegel et al. (1998) overpredict E(B--V)
when E(B--V)$_{NED}$ $>$ 0.15 (see e.g., Arce \& Goodman 1999; Beers et al. 2002; Yasuda et
al. 2007). In this sense we adopt the correction as suggested by Bonifacio,
Monai \& Beers (2000a)\footnote{E(B--V) = 0.10 + 0.65 × [E(B--V)$_{NED}$ -- 0.1]}.
On the other hand for those stars in the dust layer the colour excess along the line of sight must
be estimated assuming that the dust in the Galactic disk can be modelled as a thin
exponential disk with a scale-height of 125 pc (Bonifacio et al. 2000b; Beers et al. 2002), so

\begin{equation}\label{eq1}
E(B-V) = [ 1 - \exp(-|d \sin(b)|/h]~ E(B-V)_{NED}, 
\end{equation}

\noindent where E(B--V) is the full colour excess for the corresponding star at
the distance {\it d}, {\it b} is the galactic latitude and {\it h} is the scale-height of the
thin dust disk which adopted 125 pc. For those objects with high and intermediate galactic
latitude ({\it b} $>$ $\pm$10$^{0}$) we adopt the values calculated by the NED due to
the interstellar dust in the line of sight is reduced significantly. A distribution of the
total colour excesses found for our sample stars is shown in Figure~\ref{fig2}. 
In this Figure we observe that the total reddening is small for most of the stars studied. 
This suggests that mostly the redness of the sample come mainly from 
interstellar component.

In order to check the quality of our E(B--V) estimated from the map of Schlegel et al. (1998), we compare
it with the E(B--V) calculated by Luna et al. (2008), which used the Galactic 3D-extinction
model map by Drimmel et al. (2003). Both E(B--V) values, named above, reflect only the interstellar
component in the light of sight. Table~\ref{table2} shows the number of stars, the
total colour excess observed (taken from Col. 2 of the Table 2 from Luna et al.), the
maximum colour excess of the interstellar component derived from Drimmel et al.'s map 
(taken from Col. 7 of the Table 4 from Luna et al.), the colour excess of the interstellar component
obtained from Schlegel et al.'s map (taken from Col. 5 of the Table 2), and the minimum
colour excess of the circumstellar component (taken from Col. 8 of the Table 4 from Luna et al.).
There are only seven common objects in both samples.

Figure~\ref{fig3}~shows the correlation between the colour excess derived from both maps
(upper panel) and the E(B--V) residuals as a function of colour excess derived from 
Schlegel et al.'s map, and the intrinsic colour (middle and bottom panel). 
In the upper panel we see that there are not systematic differences between the two 
samples and their residuals do not show dependence neither the
colour excess nor the intrinsic colour. Our measurements the colour excess are obtained
with a mean error of about $\pm$ 0.10.

To determine which of the two components (interstellar or circumstellar) dominates the 
observed extinction we follow criteria used by Luna et al. (2008). According to
Luna et al. (2008) in order to disentangling interstellar and circumstellar extinction
in a given source is necessary a statistical approach since it is very difficult to rely only 
on the available observations. For this we represent E(B-V)$_{NED}$ vs.
the galactic distribution of our post-AGB stars and compared them with the sample taken from catalogue 
of Guarinos (Guarinos 1988a,b; 1997). For further details of this procedure see section 4.3 
and Figure~5 in Luna et al. (2008). Only a very small group of objects in the sample
present a slight contribution of the circumstellar component. This small group  
(HD 107369, HD 112374 and 161796) is located at galactic latitude b $>$ $+$29 degrees.
At this latitude, the interstellar dust is scarce and the colour excess for these 
objects are 0.079, 0.079 and 0.030, respectively. This suggests that the contribution 
circumstellar of these objects is very low.

However, it is very likely that some objects that have a moderate (although possibly
significant) contribution from the circumstellar shell to the observed extinction may have
escaped to detection or vice-versa (Luna et al. 2008). For example, the case of AR Pup star 
in which has been detected a large contribution of circumstellar material or BD$+$39$^{o}$4926
where was no IR excess detected by IRAS (de Ruyter et al. 2006 and internal references).

In Table~\ref{table3} we represent the E(B--V) due only to the contribution of 
interstellar dust along the line of sight. With these values of E(B--V) we can de-redden 
the infrared photometry using the equations of Fiorucci \& Munari (2003) for 2MASS photometry. 
These equations are the following:

\begin{eqnarray}\label{eq2}
J_{0} = J - 0.887~E(B-V),\nonumber \\
(J-H)_{0} = (J-H) - 0.322~E(B-V),\nonumber \\
(H-K_{s})_{0} = (H-K_{s}) - 0.183~E(B-V).
\end{eqnarray}

\noindent Table~\ref{table3} shows the apparent magnitudes and the
intrinsic colours for each of our calibrator stars.
The mean errors of the observed colours are about 0.114 mag and 0.159 mag
in both colours respectively. These were obtained from
the correlation between the errors of the observed colours against the intrinsic
colours in the near-infrared passband (see Bilir et al. 2008a; Yaz et al. 2010).

\begin{figure}
\begin{center}
\includegraphics[width=7.5cm,height=7.5cm]{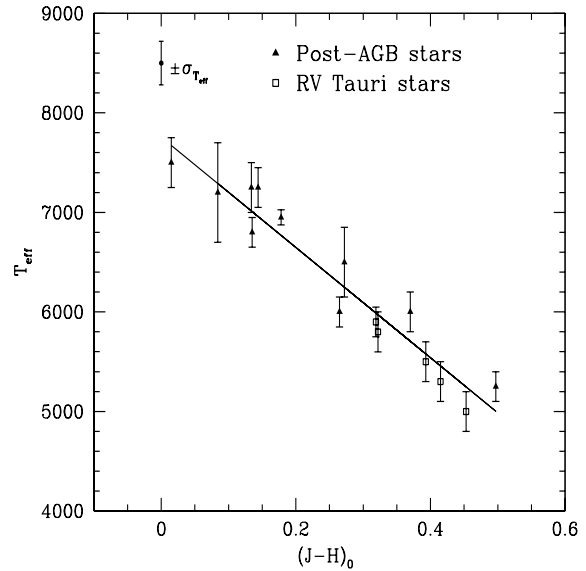}
\caption{Representation of the effective temperature against the
intrinsic colour. The straight line represent the best fit to the data.
The filled triangles represent the post-AGB stars and open squares to
the RV Tau stars. The error bars come from the uncertainties in
the sample and the T$_{\rm eff}$ relationship.}
    \label{fig4}
\end{center}
\end{figure}

\begin{figure}
\begin{center}
\includegraphics[width=7.5cm,height=7.5cm]{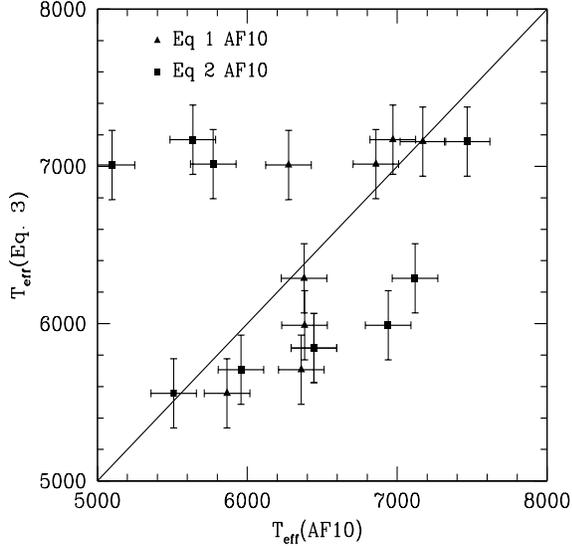}
\caption{Representation of the calculated effective temperature estimated by the
Eq.~(\ref{eq3}) against the effective temperature of Equations (1) and (2)
from Arellano Ferro (2010). The straight line represent a slope to 45 degree.
The error bars come from the uncertainties in the sample and the T$_{\rm eff}$ relationships.}
    \label{fig5}
\end{center}
\end{figure}

\begin{figure}
\begin{center}
\includegraphics[width=7.5cm,height=7.5cm]{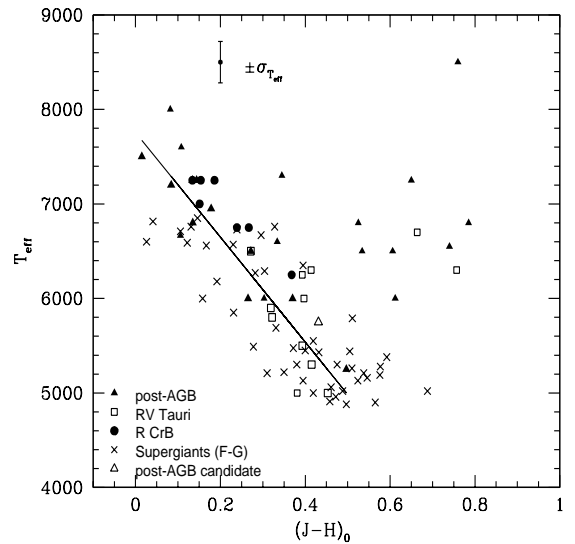}
\caption{Representation of some young and evolved sources relative to
the T$_{\rm eff}$ vs. (J--H)$_{0}$ plane. The crosses correspond to variable, double
and no-variable young supergiant stars. Filled circles represent R CrB stars.
The empty triangle correspond a one post-AGB candidate star. Filled
triangles and empty squares represent the post-AGB and RV Tauri of the
sample studied. The Eq.~(\ref{eq3}) which is represented as a solid line.
The error bar correspond to a typical uncertainty of $\pm$220~K
in T$_{\rm eff}$.}
\label{fig6}
\end{center}
\end{figure}

\section{Atmospheric parameters}
\label{sec:atmoph}

In this section, we present a set of functional relationships that allow us to derive
the atmospheric parameters using the intrinsic colours as independient variables at the
infrared region. We use the infrared region because of the infrared colours are less subject
to the effect to interstellar absortion.

When the number of independent variables is greater than one we use the method adopted
by Stock \& Stock (1999). This method developed a quantitative method to obtain
stellar physical parameters such as absolute magnitude, intrinsic colour, and a
metallicity index using the pseudo-equivalent
widths of absortion features in stellar spectra by means of polynomials and a
consistent algorithm (Molina \& Stock 2004).

\subsection{Effective temperature}
\label{sec:effective}

In order to obtain the effective temperature we use a functional
relationship involving only the intrinsic colour (J--H)$_{0}$. This relationship
has the form

\begin{eqnarray}\label{eq3}
T_{\rm eff} &=& (7756 \pm 126) - (5539 \pm 419) (J-H)_{0}, 
\end{eqnarray}

\noindent where this new relationship is valid for a range in effective temperature between
5000 $\leq$ T$_{\rm eff}$ $<$ 8000 K and intrinsic colour between 0.00 $\leq$
(J--H)$_{0}$ $\leq$ 0.50, respectively. The correlation between effective temperature and
intrinsic colour can be seen in Figure~\ref{fig4}. This dependence has been observed by
Straizys \& Lazauskaite (2009) for a large sample of dwarf and giant stars.
The standard deviation derived from the Eq.~(\ref{eq3}) is $\pm$220 K.

To verify the values of the effective temperature with
an independent method, we compare our values with the calculated effective temperature
of the functional relationships (1) and (2) from the work of Arellano Ferro (2010).
Figure~\ref{fig5} shows the relationship between the effective temperature obtained by the
Eq.~(\ref{eq3}) of this work
 with respect to the effective temperature derived from Equations (1) and
(2) of Arellano Ferro (2010). The error bars show the uncertainties of the functional
relationships of both studies. We notice that most of the stars seem to fit with a great
dispersion to the slope to 45 degrees.

In Figure~\ref{fig5} we observe that for temperature below 6500~K, the effective temperature
calculated from the relationships of Arellano Ferro (2010) appear to be relatively higher
than that calculated from Eq.~(\ref{eq3}). On the 
contrary for temperature $\sim$7000~K the values are more consistent except for three
stars HD 46703, HD 56126 and HD 179821 which the Eq. (2) of Arellano Ferro (2010)
generates outliers. The temperature values derived from the index [$c_{1}$] (filled triangles) 
seem to fit better to values of temperature calculated from Eq.~(\ref{eq3}).

In Figure~\ref{fig6} we represent the Eq.~(\ref{eq3}) and therefore,
we analyze the behavior of some young and evolved objects 
within of the effective temperature versus intrinsic colour plane. The total sample includes 
young supergiant stars (crosses) obtained from the sample of Arellano Ferro (2010) 
which have a one post-AGB candidate (empty triangle) star. A set of 7 R CrB stars (filled circles) 
taken from Stasi\'nska et al. (2006) and the total sample studied of post-AGB (filled triangles)
and RV Tauri (empty squares) stars are also included.
We observe within the studied range of colours that young supergiant stars show a 
similar trend but with a large dispersion relative to post-AGB and RV Tauri stars.
R CrB stars continue the tendency within the uncertainty of 
Eq.~(\ref{eq3}) just like the post-AGB candidate star by their nature post-AGB.
However, a large group of objects has a large dispersion and shows no-correlation with 
Eq.~(\ref{eq3}).

A possible explanation for the scattering of these objects may come from the poor quality of
observed colours, to large erros in the estimated colour excess,
or due to a significant contribution of the circumstellar component  
not detected on these objects: HD 95767, HD 101584, HD 108015, HD 163506, HD 172481
IRAS 05341+0852, IRAS 07430+1115, IRAS 19386+0155, IRAS 18095+2704, HD 82084 and AR Pup.
Luna et al. (2008) have pointed out that the reddening of the post-AGB stars of high 
galactic latitude comes mainly from circumstellar shell not detected in the line of sight.
In fact, De Ruyter et al. (2006) has found Keplerian discs of circumstellar 
dust around post-AGB and RV Tauri stars as HD 95767, HD 108015, HD 163506, HD 82084
and AR Pup. This confirms the hypothesis that most of the redness of these objects come from 
circumstellar material.

For the purpose of improving the calibration of the effective temperature from Eq.~(\ref{eq3}),
taking into account what was said above, we suggest two criteria that allow us to restrict 
those objects evolved primarily dominated from the circumstellar component.  
First, we should consider the range of validity of Eq.~(\ref{eq3}), i.e., 5000 $\leq$ 
T$_{\rm eff}$ $<$ 8000 K 
and  0.00 $\leq$ (J--H)$_{0}$ $\leq$ 0.50, respectively. Second, we should measure the colour excess  
exclusively dominated by the interstellar component. 
The mean error in the colour excess of $\pm$0.10 leads to an uncertainty of about 160 K, 
this indicates that the greater weight on the uncertainty of the effective temperature 
come from the colour excess.

\begin{figure}
\begin{center}
\includegraphics[width=7.5cm,height=7.5cm]{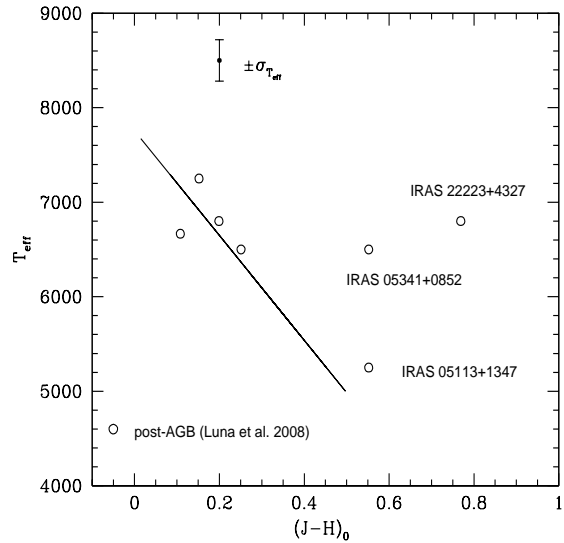}
\caption{Representation of seven common objects in
the T$_{\rm eff}$ vs. (J--H)$_{0}$ plane. The empty circles correspond to the sample in 
Luna et al. (2008). The Eq.~(\ref{eq3}) is represented as a solid line.
The error bar correspond to a typical uncertainty of $\pm$220~K
in T$_{\rm eff}$.}
\label{fig7}
\end{center}
\end{figure}

Qualitatively this procedure can be seen in Figure~\ref{fig7}. This Figure shows the seven common objects 
which are recorded in Table~\ref{table2}. The Eq.~(\ref{eq3}) which is 
represented as a solid line, the error bar corresponds to a typical uncertainty of $\pm$220~K
in T$_{\rm eff}$, and the empty circles correspond to objects evolved taken from Luna et al. (2008).
The intrinsic colour is dominated by the interstellar component derived from Drimmel et al.'s
map. We notice that objects with little or nothing of colour excess can be fit to 
Eq.~(\ref{eq3}). On the contrary, the objects exclusively dominated by the circumstellar
component (IRAS 05113+1347, IRAS 05341+0852 and IRAS 22223+4327) are clearly away from the calibration.
This confirms the validity of Eq.~(\ref{eq3}) for evolved objects dominated by interstellar components. 

\subsection{Surface gravity}
\label{sec:gravity}

In order to calculate the gravity we find a linear correlation of the gravity
as a function of intrinsic colours (J--H)$_{0}$ and (H--K$_{s}$)$_{0}$. This
has the following form:

\begin{eqnarray}\label{eq4}\nonumber
log~g &=& (0.78 \pm 0.09) - (0.55 \pm 0.31) (J-H)_{0} \\
&+& (0.90 \pm 0.19) (H-K_{s})_{0}.
\end{eqnarray}

\noindent This relationship is valid for a range of colours between
0.024 $\leq$ (J--H)$_{0}$ $\leq$ 1.249 and 0.062 $\leq$
(H--K$_{s}$)$_{0}$ $\leq$ 1.431, respectively.
The standard deviation of this relationship is $\pm$0.27.

\begin{figure}
\begin{center}
\includegraphics[width=7.5cm,height=7.5cm]{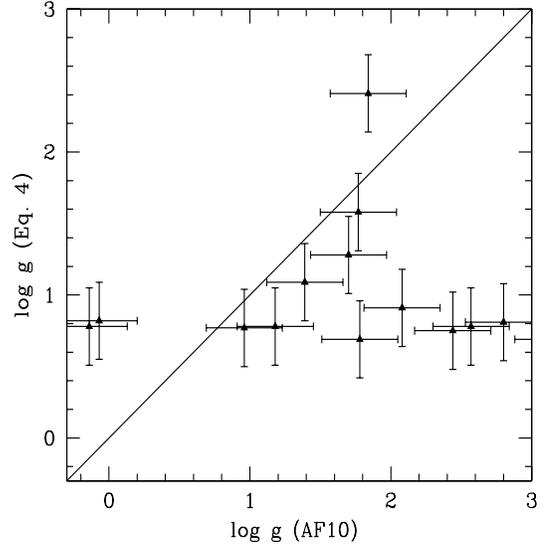}
\caption{Representation of the calculated surface gravity by the Eq. (4) as a function
of surface gravity derived by the Eq. (5) from Arellano Ferro (2010).
The comment that follows is similar to Figure~\ref{fig5}.}
\label{fig8}
\end{center}
\end{figure}

The gravities derived from Eq.~(\ref{eq4}) in this work are compared
with the gravities obtained by Arellano Ferro (2010).
In Figure~\ref{fig8} we plot the
calculated gravity by the Eq.~(\ref{eq4}) as a function of the gravity
obtained by the Eq. (5) from Arellano Ferro (2010).

\begin{table*}
 \centering
 \begin{minipage}{140mm}
  \caption{Fundamental data used in the determination of the absolute magnitude in the
near-infrared band.}
\label {table4}
\begin{tabular}{lccccl}
  \hline
\multicolumn{1}{l}{Star}&
\multicolumn{1}{c}{$\pi$}&
\multicolumn{1}{c}{$\Delta$$\pi$}&
\multicolumn{1}{c}{{\it d}}&
\multicolumn{1}{c}{E(B--V)}&
\multicolumn{1}{r}{{\it $M_{J}$}}\\
\multicolumn{1}{l}{}&
\multicolumn{1}{c}{(mas)}&
\multicolumn{1}{c}{(mas)}&
\multicolumn{1}{c}{(pc)}&
\multicolumn{1}{c}{}&
\multicolumn{1}{r}{(mag)}\\

 \hline
HD 27381     & 1.03 & 0.98 &  971 & 0.797 & $-5.62$ \\
HD 46703     & 0.89 & 1.02 & 1124 & 0.085 & $-2.52$ \\
HD 101584    & 1.35 & 0.64 &  741 & 0.150 & $-3.53$ \\
HD 105578    & 0.61 & 1.13 & 1639 & 0.098 & $-3.54$ \\
HD 114855    & 2.32 & 1.06 &  431 & 0.116 & $-1.76$ \\
HD 148743    & 0.73 & 0.43 & 1370 & 0.213 & $-5.36$ \\
HD 163506    & 0.76 & 0.23 & 1316 & 0.058 & $-5.65$ \\
HD 170756    & 0.53 & 1.40 & 1887 & 0.128 & $-5.79$ \\
HD 179821    & 0.18 & 1.12 & 5556 & 0.739 & $-9.01$ \\
HD 190390    & 1.38 & 0.42 &  725 & 0.113 & $-4.39$ \\
AR Pup       & 2.21 & 1.23 &  452 & 0.974 & $-1.25$ \\

\hline
\end{tabular}
\end{minipage}
\end{table*}

In Figure~\ref{fig8} we see that some of the gravities calculated from Eq. (5)
generate lowest and highest values of gravity for the range studied (0.0 - 2.0). 
This may be a consequence that the functional relationship derived
from the plane log~$g$ -- $\Delta$[$c_{1}$], is valid only for a few RV Tauri
($\sim$5 stars). It is
not the case for post-AGB stars of the sample which follows a different trend than 
those of young
supergiants and RV Tauri stars (Arellano Ferro 2010).

\section{Absolute magnitude}
\label{sec:magnitude}

One of the problems affecting the absolute magnitude in the post-AGB stars
is the lack of precise measures of their parallaxes, (e.g., negative parallaxes
or very large errors in their parallaxes). Our sample calibrator post-AGB stars and
RV Tau stars went through this situation, and even more, in the case of RV Tauri stars
no parallax determination existed for some of them.

The interstellar extinction for the stars in the direction of the galactic plane
is another problem when we estimate the absolute magnitude. We use the corrected
parallaxes of van Leeuwen (2007) to calculate the absolute magnitudes for a group
of stars (Table~\ref{table4}). The absolute magnitude of the stars in the
Table~\ref{table4} have been computed by the well-known distance-modulus formula
in the near-infrared $M_{J}$ band:

\begin{equation}\label{eq5}
M_{J} = J_{0} - 5 \log(1/\pi) + 5.
\end{equation}

In the estimation of the absolute magnitude we propose a functional relationship where
$M_{J}$ is a function of intrinsic colours of the following form:

\begin{eqnarray}\label{eq6}\nonumber
M_{J} &=& -(6.00 \pm 0.35) + (11.47 \pm 4.30) (J-H)_{0} \\ \nonumber
&-& (17.00 \pm 3.13) (J-H)_{0}^{2} + (3.07 \pm 0.39) \\
& & (H-K_{s})_{0}.
\end{eqnarray}

\noindent The standard deviation of the absolute magnitude in the near-infrared
band is about 0.28 mag. This value would lead to a parallax error ($\sigma_{\pi}$/$\pi$)
of about 20\% or less. The Eq.~(\ref{eq6}) is valid for a range of colours between
0.015 $\leq$ (J--H)$_{0}$ $\leq$ 0.761 and 0.046 $\leq$
(H--K$_{s}$)$_{0}$ $\leq$ 1.357, respectively.
The near-infrared region promises a better estimation
of the absolute magnitude for post-AGB and RV Tauri stars. In fact,
Bilir et al. (2008b) has observed the sensitivity of the intrinsic colours in the
near-infrared band to the absolute magnitude to a group of 44 detached binaries.
Table~\ref{table4} shows the parallax $\pi$, the parallax error $\Delta$$\pi$, the distance {\it d},
the interstellar extinction E(B--V) and the absolute magnitude $M_{J}$ in the
near-infrared band. We can observed that the large errors on the parallaxes and the colour excesses
leads to a large
error for the distance and absolute magnitude determinations, i.e. HD 179821.

\begin{figure}
\begin{center}
\includegraphics[width=7.5cm,height=7.5cm]{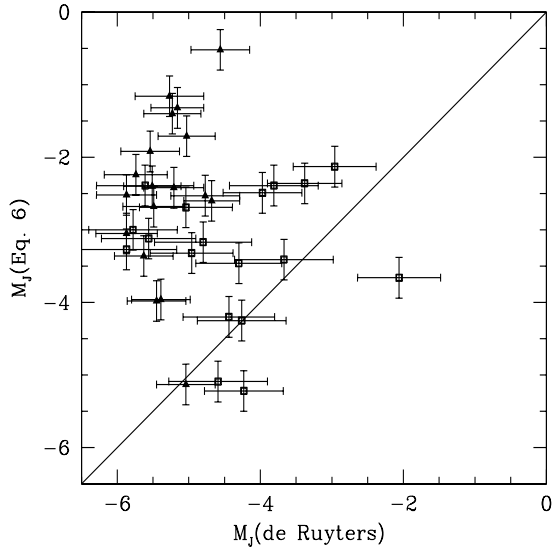}
\caption{Absolute magnitudes predicted by the equation~(\ref{eq6}) versus near-infrared
absolute magnitudes obtained of distances from de Ruyter et al. (2006).
The empty squares represent the absolute magnitudes for stars that have a defined
period P while filled triangles represent those stars without determined pulsation period P.}
\label{fig9}
\end{center}
\end{figure}

Figure~\ref{fig9} shows the estimated absolute magnitude of the
Eq.~(\ref{eq6}) against the absolute magnitude calculated by two differents methods
from de Ruyter et al.
(2006).

De Ruyter et al. (2006) carried out a homogeneous and systematic study of the Spectral
Energy Distributions (SEDs) of a sample of post-AGB and RV Tauri objects, assuming the
circumstellar dust is stored in Keplerian rotanting passive discs. Some of these
objects are considered post-AGB binary stars and also include new RV Tauri star
candidates. In determining the distance the authors used the relationship P--L from
Alcock et al. (1998) derived for a set of RV Tauri stars in the Large Magellanic
Cloud (LMC) with a defined period of pulsation. For those objects that do not have a
defined period of pulsation, Ruyter et al. (2006) took a constant relationship in determining
the luminosity,\footnote{L = (5000 $\pm$ 2000)L$_{\odot}$} assuming that the typical
luminosity of a lower mass post-AGB star is expected to be between 1000 and
10~000L$_{\odot}$. Consequently, the uncertainties for the luminosity and the
distances obtained by both methods are significant. The results of these distances
and uncertainties are shown in their Ruyter et al., Table A.2.

On the other hand, to verify the level of reliability of Eq.~(\ref{eq6}) we apply our method to
the sample of post-AGB and RV Tauri studied by the de Ruyter et al. (2006). This procedure takes
place because our sample has very few common objects in relation to the sample from de Ruyter
et al. (2006). We studied about 36
stars of a total sample of 51 post-AGB and RV Tauri stars contained within the A--G
spectral range. We identify the JHK$_{s}$ colours observed from the 2MASS source and it was found the
intrinsic colours (J--H)$_{0}$ and (H--K)$_{0}$ from the excess color derived for the sample
(see Ruyter et al., Table A.1). Finally, the absolute magnitudes are derived from Equations (5) and (6).

In Figure~\ref{fig9} we observe that a majority of the sample objects with a known period P
(open squares), the absolute magnitude $M_{J}$ derived of the distance has a tendency
to adjust to the results obtained from our functional relationship within $M_{J}$
$\leq$ $-$2 mag. This indicates that the Eq.~(\ref{eq6}) can be applied to binary or
variables objects.
However, most of the objects without pulsation period (filled triangle) and whose
distances were derived assuming a typical luminosity of a post-AGB star of low-mass,
the absolute magnitudes tend to be larger than the absolute magnitudes derived from
the Eq.~(\ref{eq6}) and are clustered in the range between 0 $\leq$ $M_{J}$ $\leq$ $-$4 mag.
The systematic differences observed for the latter group may be probably due
to differences in the masses of these objects.

\section{Discussion and Conclusion}
\label{sec:result}

The study of a sample of post-AGB and RV Tauri stars
has led to find out a set of empirical relationships that allow us to estimate the
physical parameters as the effective temperature and the gravity with precision
of 220~K and 0.27, respectively. 
Effects of variability and pulsation in RV Tauri and post-AGB stars affect 
very little to temperature and gravity calibrations. In fact, few of them
are out of calibrations, e.g., in Teff, HD 82084, AR Pup, HD 108015, HD 163506, and HD 172481.
These uncertainties derived from our effective temperature and gravity calibrations are
comparable with those derived by Arellano Ferro (2010) who used {\it ubvy$\beta$}-photometry.
These equations cover a wide range of colours in the near-infrared band and it can be valid for
variables, no-variables and/or semi-regular pulsanting objects with low-interstellar extinction
and low-masses. Furthermore, these functional relationships can be extended to R CrB stars with
a rare H-deficient. In addition, for young supergiant stars it can not be deduced their
atmospheric parameters with high precision.

For those very reddening objects whose central star is visible in the optical it is possible
to obtain a better estimation of their temperature and gravity
if we study the diffuse bands of the circumstellar material
around these stars in order to correct the effects of circumstellar reddening  
according to the procedure by Luna et al. (2008).
 
In Figure~\ref{fig2}, we also observe that the colour excess of most stars due to interstellar
dust is small, so that the reddening present in some post-AGB and RV Tauri objects such
as HD 95767, HD 101584, HD 108015, HD 163506, HD 172481, IRAS 05341+0852, IRAS 07430+1115, 
IRAS 19386+0155, IRAS 18095+2704, HD 82084 and AR Pup comes likely from circumstellar dust 
not detected stored in Keplerian discs around them (de Ruyter et al. 2006). On the contrary,
The effect of circumstellar reddening observed in the post-AGB stars such as HD 107369, HD 112374
and HD 161769 appears to be minimal since it does not affect the de-reddened colours.

On the other hand, the absolute magnitude in the near-infrared J-band promises to be
recovered satisfactorily. In fact, the magnitude absolutes of RV Tauri stars obtained from the P--L
relation are consistent within a large dispersion to the absolute magnitudes estimated in this work.
This indicates that the P--L relation could be also applicable to field stars.
The Hipparcos parallax could be the main cause contributing to the generation of the
dispersion in Eq.~(\ref{eq6}), but there are also other uncertainties that are present as the
errors in the excess of colour and the intrinsic colour. Similarly, the propagation of errors 
from the P--L
relation leads to uncertainties in the luminosities and hence, in distances. Finally,
the dust surrounding the star seen nearly edge-on also affects the luminosity
when using the P--L relation. 

We notice that for stars without defined pulsation period, the approach
all the stars may have the same luminosity is not valid because, as shown in Figure~\ref{fig9}
absolute magnitudes derived from the Eq.~(\ref{eq6}) vary between a range from
0 $\leq$ $M_{J}$ $\leq$ $-$4 mag. This indicates that this group of stars do not
have neither the same stellar masses nor an unique value of luminosity (L = 5000L$_{\odot}$).

In short, an alternative way to achieve real improvements in the estimation of the absolute magnitude
could be the use of high resolution spectroscopic techniques (see Arellano Ferro 2003) or
expected precise measurements of parallaxes in the future through the GAIA mission.

\section*{Acknowledgments}

This work has made extensive use of SIMBAD database and the ADS-NASA to which we are
thankful. I would also like to thank to Dr. Anibal Garc\1a Hern\'andez and Dr. Jes\'us Hern\'andez
for very useful and fruitful comments and suggestions on the manuscript. The author thank to an
anonymous referee for helpful suggestions. I would also like to thank Prof. Luz Angela Ca\~nas for 
her valuable English corrections.

\end{document}